\documentclass[10pt,twocolumn,twoside]{pnas-new}
\usepackage[version=4]{mhchem}
\usepackage{xspace}
\DeclareSymbolFont{UPM}{U}{eur}{m}{n}
\DeclareMathSymbol{\umu}{0}{UPM}{"16}
\newcommand\micro{$\umu$}
\newcommand\micron{\micro m\xspace}
\newcommand\microns{\micro m\xspace}

\templatetype{pnasresearcharticle} % Choose template
\begin{document}

\title{A first look at rocky exoplanets with JWST}

% Use letters for affiliations, numbers to show equal authorship (if applicable) and to indicate the corresponding author
\author[a,1]{Laura Kreidberg}
\author[b,c,1]{Kevin B. Stevenson}

\affil[a]{Max Planck Institute for Astronomy, K\"onigstuhl 17, 69117 Heidelberg, Germany}
\affil[b]{Johns Hopkins University Applied Physics Laboratory, Laurel, MD, 20723, USA}
\affil[c]{Consortium on Habitability and Atmospheres of M-dwarf Planets (CHAMPs), Laurel, MD, 20723, USA}

% Please give the surname of the lead author for the running footer
\leadauthor{Kreidberg}

% Please add a significance statement to explain the relevance of your work

\significancestatement{Rocky planets are very common around other stars, with tens of billions predicted to exist in our Galaxy alone. Until recently, however, it was not possible to characterize these planets in detail, or learn about their atmospheric properties. Thanks to the new observing capabilities of the James Webb Space Telescope, the first constraints on realistic atmospheres for rocky exoplanets are now possible. The first data are beginning to constrain the inventory of volatiles (carbon, nitrogen, and oxygen-bearing molecules), providing important context for the eventual search for life on other worlds.}

% Please include corresponding author, author contribution and author declaration information
\authorcontributions{L.K. led the overall manuscript and focused on the thermal emission section. K.B.S. led the transmission spectroscopy section. Both authors contributed text and figures.}
\correspondingauthor{\textsuperscript{2}To whom correspondence should be addressed. E-mail: kreidberg@mpia.de}

% At least three keywords are required at submission. Please provide three to five keywords, separated by the pipe symbol.

\begin{abstract}
Rocky exoplanet characterization has been a top priority for early James Webb Space Telescope (JWST) science operations.  Several milestones have been achieved, including the most precise rocky planet transmission spectra measured to date, and the first detection of thermal emission for rocky worlds below 800 Kelvin. Despite these advances, no atmospheres have been definitively detected. Several transmission spectra show tentative evidence for molecular absorption features, but these hints are marginally significant and the spectra may be affected by stellar contamination. Features from many plausible atmospheres, including those dominated by oxygen, nitrogen, and carbon dioxide, are below the current noise level.  Meanwhile, the emerging picture from thermal emission spectra is that the planets have hot daysides, consistent with either a bare rock composition or low surface pressure atmospheres ($<10$ bar). Higher surface pressures and high carbon dioxide abundances are generally ruled out, assuming cloud-free atmosphere models. The absence of strong \ce{CO2} features hints at a limited initial volatile inventory or rapid atmospheric escape during the planets' early lifetimes. Taken together, these results motivate a push towards higher precision data, as well as observations of cooler planets that may be more likely to retain atmospheres.  As a goal for future transmission spectroscopy, we suggest a ``five scale height challenge," to achieve the precision necessary to detect \ce{CO2} features in nitrogen-rich atmospheres.  Detecting rocky planet atmospheres remains challenging, but with JWST's excellent performance and a continuing investment of telescope time, we are optimistic these uncharted atmospheres will be detected in coming years.

%Please provide an abstract of no more than 250 words in a single paragraph. Abstracts should explain to the general reader the major contributions of the article. References in the abstract must be cited in full within the abstract itself and cited in the text.
% James Webb Space Telescope (JWST)
\end{abstract}

\dates{This manuscript was compiled on \today}
\doi{\url{www.pnas.org/cgi/doi/10.1073/pnas.XXXXXXXXXX}}

\maketitle
\thispagestyle{firststyle}
\ifthenelse{\boolean{shortarticle}}{\ifthenelse{\boolean{singlecolumn}}{\abscontentformatted}{\abscontent}}{}

\firstpage{3}
% Use \firstpage to indicate which paragraph and line will start the second page and subsequent formatting. In this example, there are a total of 11 paragraphs on the first page, counting the first level heading as a paragraph. The value {12} represents the number of the paragraph starting the second page. If a paragraph runs over onto the second page, include a bracket with the paragraph line number starting the second page, followed by the paragraph number in curly brackets, e.g. "\firstpage[4]{11}".

% If your first paragraph (i.e. with the \dropcap) contains a list environment (quote, quotation, theorem, definition, enumerate, itemize...), the line after the list may have some extra indentation. If this is the case, add \parshape=0 to the end of the list environment.

\dropcap{S}ince the first discoveries of rocky planets orbiting other stars, anticipation has grown regarding the nature of their atmospheres \citep{dressing15, petigura13}. However, very little can be determined \emph{a priori} about their properties -- even whether an atmosphere is present! In contrast to large gas giants, small rocky worlds have interiors made primarily of iron and rock.  Any atmospheric layer on top is so thin that it cannot be inferred via mass and radius measurements alone \citep{dorn15}. Direct, spectroscopic measurements of an atmosphere are necessary to confirm its existence and determine fundamental properties such as surface pressure and chemical composition.  For this article, we use mass and radius measurements as a starting point to identify rocky planets. We consider a planet to be rocky if its density is consistent with a pure silicate composition or higher, and also discuss a few borderline cases to provide context. We assume that any atmosphere present has high mean molecular weight (with negligible hydrogen and helium). We adopt 10 bars as the boundary between thin and thick atmospheres, marking the threshold at which full day-to-night heat redistribution is expected \citep{koll22}.

 %A  density consistent with a rocky composition suggests a solid surface, and most planets smaller than 1.6 Earth radii are likely rocky \citep[e.g.,][]{rogers15}, though there are some exceptions \citep{demangeon21,piaulet23}. The exact transition between rocky and gaseous planets depends on the nature of the atmosphere.

Prior to the launch of JWST, there were limited observational constraints on rocky exoplanet atmospheres. The primary finding was that cloud-free, hydrogen-rich compositions were ruled out for a sample of nine planets, based on featureless transmission spectra \citep{dewit16,dewit18, diamondlowe18, wakeford19, diamondlowe20a, diamondlowe20b, edwards21, libbyroberts22, Garcia2022}.  
%Zhanbo Zhang + 2018
For a few hot planets, thick atmospheres  were disfavored based on Spitzer Space Telescope observations \citep{kreidberg19,zieba22,crossfield22}. There was some evidence for atmospheric circulation from the thermal phase curve of the lava planet 55 Cancri e, but this was not confirmed by subsequent analysis \citep{demory16,angelo2017,hammond17,mercier22}.  Overall, these results match theoretical predictions that hydrogen-rich atmospheres are unlikely for hot rocky exoplanets, as this scenario requires a finely-tuned production and loss rate for hydrogen \citep{hu23}. However, more plausible scenarios were largely consistent with the data, including high mean molecular weight atmospheres, as well as bare rocky surfaces. 

In addition to these first observations, there has also been extensive theoretical work to predict the atmospheric properties of rocky planets, particularly those orbiting M-dwarf stars \citep[reviewed in][]{scalo07, tarter07, shields16,meadows18,wordsworth22, lichtenberg23, lichtenberg24}. This work has demonstrated that the atmospheres can be shaped by a wide range of physical processes, including initial volatile delivery, atmospheric escape, interior-atmosphere interaction, and even the presence of life. Of all these processes, atmospheric loss is an especially pressing concern for JWST observations, since it is unknown which planets have retained their atmospheres.  

\begin{figure}%[tbhp]
\centering
\includegraphics[width=1.0\linewidth]{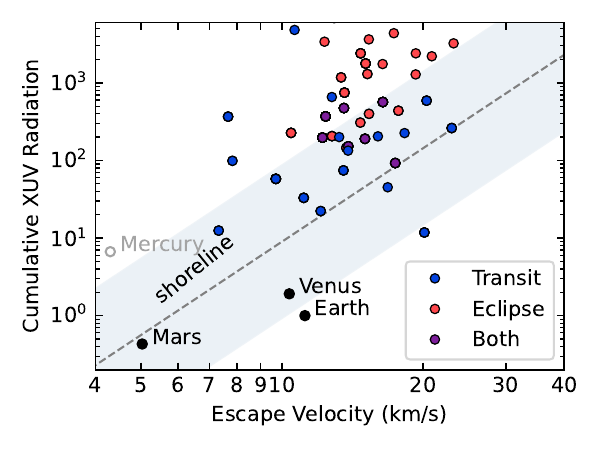}
\caption{Rocky planets with approved JWST transit and eclipse observations in Cycles 1 -- 4, relative to their escape velocity and lifetime X-ray and UV (XUV) irradiation \citep[adapated from][]{zahnle17}. Cumulative XUV radiation is normalized relative to that of the Earth.  The XUV flux is calculated for a constant system age; a reasonable assumption given that the majority of XUV flux is emitted in the first Gyr of the stars' lifetimes.  The cosmic shoreline is denoted with a gray dashed line, which divides Solar System bodies with substantial atmospheres (below the line) from those without (above). To illustrate the large uncertainties on the possible location (and even existence) of a shoreline around other stars, we shaded the region within an order of magnitude of the relation from \cite{zahnle17}.  Most planets observed by JWST are situated above the gray dashed line and, thus, are less likely to have atmospheres.
}
\label{fig:shoreline}
\end{figure}

As a guideline to determine which planets are more likely to have atmospheres, the concept of the ``cosmic shoreline" has emerged as a popular framework \citep{zahnle17}. Inspired by empirical trends in the Solar System, the shoreline is a proposed dividing line that separates bodies with atmospheres from those without (see Fig.~\ref{fig:shoreline}). The idea is that planets with a higher escape velocity and lower irradiation are more likely to retain atmospheres. However, this concept is more qualitative than quantitative, and it is still unknown how the shoreline is affected by stellar type or irradiation history. In general, though, planets at a given stellar irradiation are exposed to relatively higher levels of high-energy flux around late-type stars. This effect is especially pronounced for the latest M-dwarfs (M5 and later), which have a UV-bright phase extending as long as 6 Gyr \citep{west08}. Exposure to high-energy irradiation drives more atmospheric loss, though atmospheres are expected to survive under some conditions \citep{luger15,bolmont17,turbet18,Macdonald2019}.

With the revolutionary advance in sensitivity provided by JWST, the characterization of realistic atmospheres around rocky planets is finally possible. In particular, the combination of JWST's large aperture, stability, and near-to-mid-infrared wavelength coverage provides astronomers with their first access to the tiny signals expected from important volatile molecules expected in rocky exoplanet atmospheres, including water (\ce{H2O}), carbon dioxide (\ce{CO2}), methane (\ce{CH4}), ammonia (\ce{NH3}), carbon monoxide (\ce{CO}), and others \citep[e.g.][]{morley17b, lincowski18, ortenzi20}.  The primary methods for atmosphere characterization are transmission and emission spectroscopy. Nearly all of the feasible targets orbit small, M-dwarf host stars --- these are the most numerous rocky planets in the Galaxy, and also the easiest to characterize with transits and eclipses. Studying the atmospheres of true Earth analogs around Sun-like (FGK) stars is not possible with JWST, and will require a next-generation flagship space telescope with direct imaging capability \citep{decadal2020, quanz22}. 

% \newpage

\section{Transmission Spectroscopy}
\label{sec:transit}

\begin{figure}[t]%[tbhp]
\centering
\includegraphics[width=1.0\linewidth]{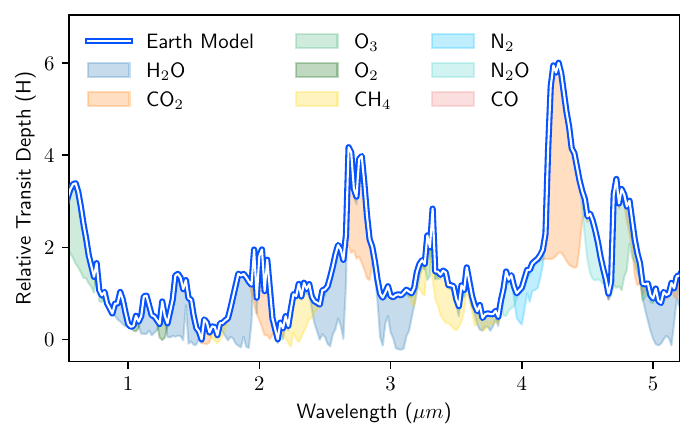}
\caption{Best-fit model to an empirical transmission spectrum of the cloudless Earth in units of scale height, $H$ (adapted from \cite{LY2023b}).  Different colors indicate the contributions from various molecules.  Despite its relatively low abundance, \ce{CO2} has the most prominent spectral features (at 2.7 and 4.3 {\microns}), making it an ideal molecule to search for in exoplanet atmospheres. The strength of different absorption peaks is sensitive to the underlying atmospheric composition and detailed atmospheric modeling can constrain the abundances of individual absorbing species (which may differ from those of the Earth). This figure neglects the effects of refraction, which effectively blocks transmission below roughly 1.5 scale heights for Earth-Sun analog systems (planets orbiting M-dwarfs are less strongly affected) \citep{betremieux14}.
}
\label{fig:earth}
\end{figure}
Transmission spectra measure the wavelength-dependent size of a planet during its transit. As illustrated in \autoref{fig:earth}, planets have a larger transit depth at wavelengths where the atmosphere is more opaque \citep{kreidberg18c}. The amplitude of spectral features is set by the atmospheric scale height ($H$), which is inversely proportional to the mean molecular weight ($\mu$) \citep{seager10}. In contrast to hydrogen-dominated gas giants with vertically extended atmospheres, rocky planets are expected to have more compact secondary atmospheres with high mean molecular weights, leading to much smaller features. JWST is the first telescope with sensitivity to detect plausible, high-mean-molecular-weight compositions using a realistic amount of observing time. %These atmospheres could range from \ce{H2O}-dominated compositions ($\mu$ = 18 g/mol) to atmospheres rich in \ce{CO2} ($\mu$ = 44 g/mol), which have roughly 2x smaller features.

To illustrate which molecules may be detectable for rocky exoplanets, \autoref{fig:earth} depicts the best-fit model to an empirical transmission spectrum of the cloudless Earth \citep{kaltenegger09, Macdonald2019, LY2023b}.  This spectrum exhibits strong absorption features from several key volatile species, including \ce{H2O}, \ce{CH4}, and \ce{CO2}. Despite its relatively low abundance, \ce{CO2} has the strongest absorption feature, spanning approximately five atmospheric scale heights at 4.3-{\micron}. This large feature has been the focus of many early JWST programs. While \ce{CH4} is less prominent than \ce{CO2} in Earth's atmosphere, its abundance may be enhanced in rocky M-dwarf exoplanet atmospheres \citep[i.e.,][]{segura05, rugheimer15, Wunderlich2019}. Similarly, ozone may be more prominent in the troposphere (via the smog mechanism) due to enhanced UV or for desiccated planets due to their large build-up of oxygen \citep[e.g.][]{Grenfell2013, luger15}.  Finally, absorption features from \ce{H2O} are also present. This molecule is particularly enticing to search for with JWST because it is a prerequisite to the formation of life as we know it; however, it also comes with the challenge that many of the planet's host stars have water vapor in their photospheres, and the signals are not trivial to disentangle (see next section).  Other molecules shown in \autoref{fig:earth} are likely undetectable with JWST.

% C1 and C2 observations
In the first two cycles of JWST observations, there were more than 60 transit observations of rocky exoplanets.  The two most frequently targeted systems are L98-59 and TRAPPIST-1.  The former has 11 successful observations across three planets and the latter has 23 successful observations split between seven planets.  Of the two dozen planets observed in transmission, $>80\%$ were targeted at least twice (usually with the same instrument mode).  The telescope performance has been excellent, with much better pointing stability and fewer near-infrared instrument systematics than seen for the Spitzer and Hubble Space Telescopes.  For bright targets, the white light curve precision is often $>2.0\times$ the predicted, photon-limited value due to the presence of red noise, but the spectroscopic light curve precision is an average of $1.3\times$ the PandExo prediction and the spectroscopic residuals are almost always dominated by Gaussian noise \citep[e.g.][]{MoranStevenson2023,Alderson2024}. When binned to a uniform resolution of 50 nm, the most precise transmission spectrum has an average uncertainty of 11 ppm (GJ~341b) and the highest SNR spectrum has an average uncertainty of $3.8$ scale heights (GJ~486b, assuming a \ce{N2}-rich atmosphere).

% FINDME: Add TRAPPIST-1c, LTT 1445Ab, LHS 1478b
% https://arxiv.org/html/2409.19333v1, https://arxiv.org/abs/2410.10987, https://arxiv.org/html/2410.11048v1
\begin{figure}[t!]%[tbhp]
\centering
\includegraphics[width=1.0\linewidth]{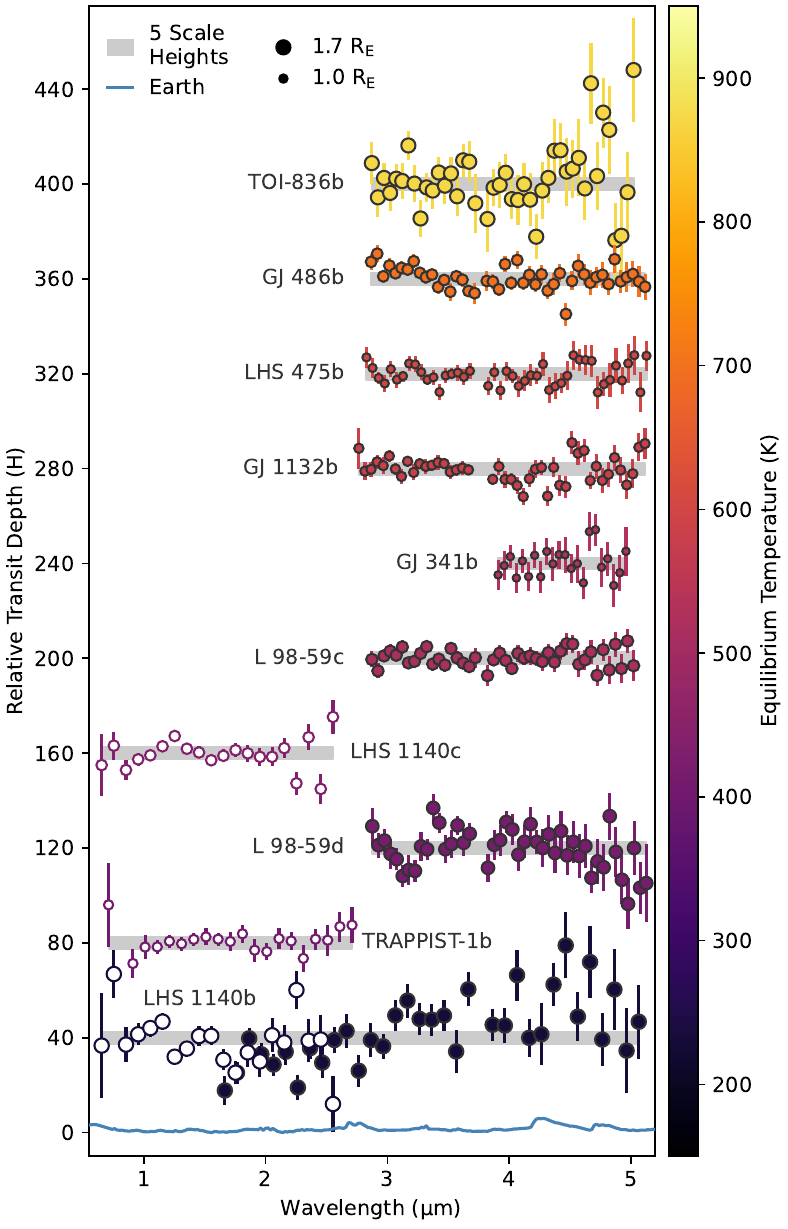}
\caption{Published transmission spectra of rocky exoplanets.  The data have been binned to a uniform resolution and converted to scale heights, $H$, assuming a pure nitrogen atmosphere ($\mu=28$). For clarity, we apply a $40H$ vertical offset between adjacent planets.  The spectra for TRAPPIST-1b and LHS 1140b have been corrected for stellar contamination. The symbol size indicates planet radius and the color indicates planet equilibrium temperature.  Open symbols depict NIRISS data and closed symbols depict NIRSpec or NIRCam data.  The gray regions, which are $5H$ in height, roughly indicate the maximum size of molecular absorption features for a nitrogen-rich atmosphere.  At the current level of precision, none of the spectra are sensitive to features in nitrogen-rich, let alone carbon-dioxide-rich, atmospheres. The blue Earth model is the same as that shown in \autoref{fig:earth}.
}
\label{fig:transits}
\end{figure}

% Level of precision varies between systems (usually with stellar mag) but normalizes pretty well when converting scale heights.
\autoref{fig:transits} shows a family portrait of currently published transmission spectra.  The planets are sorted by temperature, ranging from $T_{eq}=170-870$~K, and are normalized by their respective atmospheric scale heights assuming a pure nitrogen atmosphere ($\mu = 28$~g/mol).  Thus far, none of these planets have the requisite precision to detect an Earth-like atmosphere (shown in blue); however, the most precise transmission spectra are sensitive to cloud-free \ce{H2O}-rich atmospheres ($\mu = 18$~g/mol). 

Four planets (TOI-836b, LHS~475b, GJ~341b, and L 98-59c) have transmission spectra that are consistent with a flat line \citep{Alderson2024, LY2023a, Kirk2024, Scarsdale2024}.  The lack of spectral features makes the determination of a secondary atmosphere inconclusive because the data are consistent with a planet that has:
(1) a high-altitude cloud deck (as seen for many larger exoplanets),
(2) a high-mean-molecular-weight atmosphere (similar to Mars, Earth, or Venus), or
(3) no appreciable atmosphere (similar to Mercury).
More observations are needed to distinguish between these three scenarios.  All four planets are easily able to rule out hydrogen-rich atmospheres at high confidence.

The transmission spectra of two planets (GJ~486b and GJ~1132b) show evidence for the presence of water vapor; however, its source is unclear \citep{MoranStevenson2023, MayMacDonald2023}.  M-dwarf stars can impart spurious water features into transmission spectra due to the presence of cool, unocculted starspots.  Known as the Transit Light Source (TLS) effect, or simply stellar contamination \citep{Rackham2018,Apai2018,Barclay2021}, these features can be distinguished from those of a water-rich planetary atmosphere by acquiring high-precision, broad-wavelength (1 -- 5 {\micron}) transmission spectra.  NIRISS/SOSS observations of GJ~486b are currently planned for Cycle 3 (GO 5866, PI: MacDonald) and should resolve the water vapor degeneracy.

Of all the JWST data so far, observations of the TRAPPIST-1 system were perhaps the most hotly anticipated. The TRAPPIST-1 system consists of a late-M dwarf star with seven known transiting rocky planets, three of which are in the conservative habitable zone \cite{Kopparapu2013}.
Due to its favorable planet-to-star radius ratio, TRAPPIST-1 was initially viewed as the most promising system to yield the first detection of a secondary atmosphere.  While JWST has observed at least two transits of each planet, interpreting the data has been slowed by the TLS effect.  Contamination from unocculted stellar heterogeneities dominates the measured transmission spectra, sometimes producing signals from unocculted starspots in one transit and unocculted faculae in the subsequent transit \citep{Lim23}.  This short-term variability is consistent with a heterogeneous stellar photosphere rotating with a roughly three-day period \citep{vida17}. 

The contamination-corrected featureless spectrum of TRAPPIST-1b in \autoref{fig:transits} is consistent with the secondary eclipse interpretation in \autoref{sec:eclipse} of an airless body.   A Cycle 3 program (GO 6456; PI: Allen) is taking advantage of this fact through transit observations of planets b and e in quick succession.  The goal is to use the stellar contamination signal measured within b's spectrum to account for the contamination in planet e's spectrum.
% TRAPPIST-1h
% The contamination-corrected TRAPPIST-1h spectrum shown in \autoref{sec:eclipse} proves that the measured stellar contamination from one planet can be used to correct another planet's spectrum \citep{Rustamkulov25}.  Unfortunately, planet b's transit overlapped with that of h, thus producing larger uncertainties in the transmission spectrum than originally predicted.
 
% How does TLS depend on stellar type?

% L 98-59d sulfur
Some of the most promising evidence for spectral features thus far is the spectrum of the super-Earth, L~98-59d \citep[$\rho=0.57\rho_{\oplus}$, $R=1.58R_{\oplus}$;][]{Gressier2024}. This low-density planet is expected to have a volatile envelope \citep{demangeon21}.  Indeed, a single transit using NIRSpec/G395H favors a broad absorption feature spanning 3.3 -- 4.8 {\microns}.  Atmospheric retrievals suggest the presence of sulfur-bearing species \ce{H2S} and \ce{SO2} within a hydrogen/helium atmosphere.  While this scenario is preferred over a flat line, the exact significance varies based on the reduction pipeline, retrieval assumptions, and fitted vertical offset between the NRS1 and NRS2 detectors.  Cycle 2 observations from another program (GO 4098; PI: Benneke) should provide an important validation of L~98-59d's atmosphere.

% LHS 1140b+c
Another highly anticipated measurement was the transmission spectrum of LHS~1140b. The exact nature of this planet is uncertain.  It is a temperate world within the radius valley that could support liquid water on a rocky surface, or it could be a water world with a massive atmosphere and no well-defined surface. To distinguish between these (and other) competing scenarios, two programs each observed two transits of LHS~1140b at different wavelengths.
% Cadieux2024b:
The NIRISS data (0.6 -- 2.8 {\microns}) found strong evidence for stellar contamination and tentative evidence of a \ce{N2}-rich atmosphere with Rayleigh scattering \citep{Cadieux2024b}.
With the use of general circulation models, \cite{Cadieux2024b} are able to rule out moderately-cloudy, \ce{H2}-rich atmospheres up to $1000\times$ solar metallicity.
% Damiano2024: 
Using NIRSpec (1.7 -- 5.2 {\microns}), \cite{Damiano2024} also favor the conclusion of a \ce{N2}-rich atmosphere.  However, cloudy \ce{H2}-rich models that include the dominant carbon, nitrogen, and oxygen bearing species also achieve comparable fits (within $1\sigma$). Based on its low instellation and high surface gravity, LHS~1140b is one of the most likely transiting planets to have retained a secondary atmosphere.  Assuming that atmosphere is \ce{N2} rich and given the level of scatter seen in \autoref{fig:transits} ($11H$ with NIRISS and $14H$ with NIRSpec at $1\sigma$), LHS~1140b will require a focused observing campaign over multiple cycles to achieve the necessary precision for JWST to confidently detect its atmosphere.  Observations will begin with JWST Cycle 4 (GO 7073; PI Lustig-Yaeger).

\section{Thermal emission}
\label{sec:eclipse}

\begin{figure*}
\includegraphics[width=1.0\textwidth]{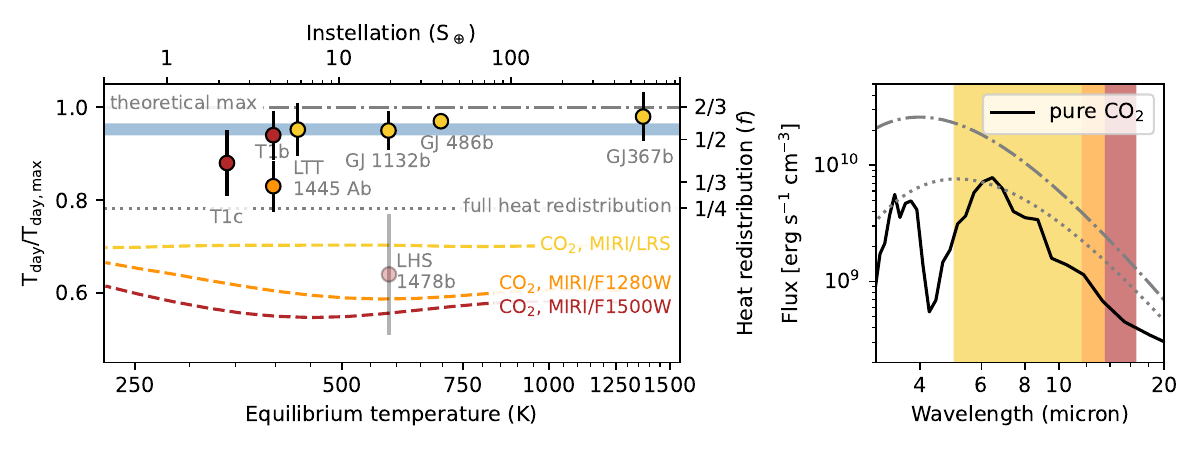}
\caption{\textbf{Left:} JWST measurements of dayside brightness temperature for rocky planets, compared to predictions from theoretical models. The values are normalized relative to the theoretical maximum dayside temperature for a bare rock at each equilibrium temperature, and the twin y axis shows the heat redistribution efficiency, $f$. The theoretical maximum corresponds to a completely absorptive black rock (zero albedo); in reality, the lowest known albedos in the Solar System are around 0.05, for space-weathered asteroids \citep[e.g.][]{shestopalov13}. The limiting cases for zero and full heat redistribution in a gray, zero-albedo atmosphere are marked with the dash-dot and dotted gray lines, respectively.  The colored lines indicate model predictions for a 10-bar, carbon dioxide atmosphere with full heat redistribution  \citep{malik19}. We show three different bandpasses, roughly 5 - 12 $\mu$m (MIRI/LRS; yellow), 11.5 - 13.5 $\mu$m (MIRI/F1280W; orange), and 13.5 - 16.6 $\mu$m (MIRI/F1500W; red). The model is coolest in the reddest filter due to the strong \ce{CO2} absorption feature centered near 15~$\mu$m. The color of the points corresponds to the bandpass used for the measurements (note that TRAPPIST-1b has a measured brightness in two different bandpasses). The blue bar indicates the $\pm1\sigma$ interval for the weighted average brightness temperature of all planets. While there is some scatter, the planets are all consistent (within 1.8$\sigma$) with a single brightness temperature equal to $0.95\pm0.01$ times the zero-albedo bare rock prediction. The data point for LHS~1478b is shown with increased transparency due to the instrument systematics that affected the data quality.  \textbf{Right:} Emitted flux versus wavelength for the three different atmosphere models, assuming a GJ-1132b-like planet ($T_\mathrm{eq} = 580$ K). The colored shading corresponds to the approximate wavelength range of the three MIRI instrument modes used for the observations (described above).%; slitless Low Resolution Spectroscopy (LRS; yellow), and two photometric filters, F1280W (orange) and F1500W (red).
}
\label{fig:emission}
\end{figure*}

Thermal emission measurements are complementary to transmission spectroscopy. To measure thermal emission, the typical approach is to observe a secondary eclipse, when the planet passes behind the star. This enables an estimate of the brightness temperature of the planet's dayside, due to the drop in flux during the eclipse. Full-orbit phase curve observations are also possible, providing a view of the thermal emission as a function of longitude. These measurements are most feasible for planets on short-period orbits, generally less than a few days. Due to strong tidal interactions with the star, the planets are expected to be in synchronous rotation, with a permanent irradiated dayside and non-irradiated nightside \citep{kasting93}. Consequently, full-orbit phase curves measure the change in brightness from dayside to nightside hemisphere.

The brightness temperature of the planet is highly sensitive to its atmospheric properties \citep{seager09, selsis11, koll16}. For an idealized gray atmosphere, the higher the surface pressure, the more heat it can transport from dayside to nightside. A bare rock absorbs stellar flux and reradiates it directly back to space, leading to a hot dayside and cold nightside. By contrast, a 100-bar Venus-like atmosphere has nearly full heat redistribution, cooling off the dayside and heating up the nightside. The dayside temperature can therefore be used as a rough proxy for the surface pressure, with models showing a rapid decrease in temperature over the pressure range of $\sim10^{-2}$ to $10^2$ bars \citep{koll19}.  This basic assessment of surface pressure is useful for rocky planets, where the presence of an atmosphere is unknown. 

The heat redistribution factor is typically parameterized as $f$, with a corresponding dayside temperature equal to $T_\mathrm{day} = (4f)^{1/4}T_\mathrm{eq}$ \citep{hansen08,koll22}. Zero redistribution (a bare rock) implies $f = 2/3$, and perfect redistribution (isotropic global temperature) gives $f=1/4$. For the purposes of this manuscript, we define a ``thick" atmosphere as having surface pressure greater than 10 bar, the pressure level where full heat redistribution is typically achieved \citep{koll22}.

In addition to the effect of atmospheric heat redistribution, the chemical composition is also crucial in shaping the emergent flux for more realistic, non-gray atmospheres \citep[e.g.][]{morley17b,lincowski18,malik19}. In the near- and mid-infrared, there are strong absorption bands from several molecules that are expected in high molecular weight atmospheres, including \ce{H2O}, \ce{CO2}, \ce{CH4}, and \ce{NH3} (see Fig. \ref{fig:earth}). The chemical composition also affects the temperature structure: absorption of stellar light in the upper atmosphere can lead to thermal inversions, where the temperature increases with altitude over some pressure range \citep{morley17b,lincowski18, malik19}. If a thermal inversion exists, spectral features are seen in emission rather than absorption.  The presence and strength of thermal inversions depend on the chemical composition, presence of aerosols, and the stellar spectrum. Finally, in addition to the chemical composition of the atmosphere, the surface composition and albedo also affect the emergent spectrum. Thermal emission spectra are, therefore, a powerful tool for measuring both the chemical composition and thermal structure of atmospheres and surfaces.

Before the launch of JWST, there were only a handful of planets smaller than $2\,R_\oplus$ with detected thermal emission from Spitzer \citep{demory16, kreidberg19, crossfield22,zieba22}. This is because the signals from temperate planets are very small and peak in the mid-infrared. The planet-to-star flux scales as $F_\mathrm{planet}/F_\mathrm{star} = (R_\mathrm{planet}/R_\mathrm{star})^2 \times \mathrm{B}(\lambda, T_\mathrm{planet})/\mathrm{B}(\lambda,T_\mathrm{star}$), where $\mathrm{B}$ is the Planck function, and $R_p/R_s$ is the planet-to-star radius ratio. With its much larger aperture and mid-infrared sensitivity, JWST is poised to increase the sample of rocky planets accessible for thermal emission measurements to several dozen, with cooler temperatures than we could detect previously. Nearly all of the feasible candidates have M-dwarf host stars. These stars are both smaller and cooler than the Sun, increasing the relative amplitude of the planet signal. JWST/MIRI is the workhorse instrument for thermal emission measurements because it provides access to mid-infrared wavelengths from $5 - 28\mu$m, where the signal-to-noise for temperate planets is largest. Still, even for the most favorable cases, the planet-to-star flux is small, at most $0.1\%$. 

The amplitude of the signal also  drops off rapidly for cooler planets, thanks to the steep $\sim T^4$ dependence of the bolometric luminosity. As an example, assuming a GJ~1132b-like planet with an albedo of 0.1, the dayside temperature is 720~K and its planet-to-star flux ratio would be 360 ppm at 15 {\microns}.  However, a similar planet in the habitable zone ($T_{day} = 285$~K) would exhibit an order of magnitude less flux (36 ppm) at the same wavelength.  This in turn requires $10\times$ higher precision to detect an atmosphere, which translates to either $100\times$ as many eclipses or $100\times$ more collecting area.  For reference, JWST has a collecting area that is $\sim45\times$ larger than that of Spitzer. Thermal emission measurements are therefore best suited for hotter targets, whereas cooler, habitable-zone planets are easier to observe with transmission spectroscopy, which is less sensitive to temperature.
% System        Tday    Fp/Fs (15 microns)
%               (K)     (ppm)
% GJ 1132       720     360
% GJ 1132       285     36

Thermal emission spectroscopy has been a priority for the first four cycles of JWST operations, with over 30 unique targets approved (see Fig.~\ref{fig:shoreline}). So far, seven planets have published thermal emission measurements: TRAPPIST-1b and c,  GJ~1132b, GJ~486b, GJ~367b, LTT~1445~Ab, and LHS~1478b \citep{greene23, zieba23, ducrot24, xue24,mansfield24,zhang24, wachiraphan24, august25}.  The measured brightness temperatures are summarized in Fig.~\ref{fig:emission}, compared to predictions for three different scenarios: a zero-albedo bare rock, a full heat redistribution gray atmosphere, and a 10 bar, pure \ce{CO2} atmosphere. The \ce{CO2} model tracks are from the grid of radiative-convective equilibrium models from \cite{malik19}, assuming full heat redistribution, no clouds, a surface albedo of 0.1, and a host star effective temperature of 3200 K. The models show strong absorption features from \ce{CO2} centered at 4.5 and 15 microns, and have cooler daysides overall than the full redistribution gray atmosphere because they have a higher Bond albedo. These 10 bar models are intended for illustrative purposes only; lower surface pressures would increase the average dayside temperature, as well as weaken the \ce{CO2} features. A wider range of thermal emission models are available in \cite{desmarais02,kaltenegger17,fujii18}.

Overall, the planets in Fig.~\ref{fig:emission} have hot daysides, close to the maximum zero-albedo bare rock temperature ($T_\mathrm{B,max}$).  The ensemble of data is consistent with a constant relative brightness temperature of $0.95\pm0.01 \times T_\mathrm{B,max}$ ($\chi^2_\nu = 1.9$; 7 degrees of freedom; $1.8\sigma$ confidence).  We note that this result is driven mainly by the hottest planets in the sample, with the most precise  temperature measurements (LTT~1445 Ab, GJ~1132b, GJ~486b, and GJ~367b). The high brightness temperatures disfavor thick atmospheres with significant heat redistribution, but are broadly consistent with either dark rocky surfaces or low surface pressures \citep{zhang24, xue24, mansfield24, wachiraphan24}. The exact constraints vary depending on the planet and the model atmospheric composition, but for simple \ce{O2}/\ce{CO2} mixtures, the data generally require less than $\sim1\%$ \ce{CO2} and less than 10 bar surface pressure. For GJ~367b, the hottest planet in the sample, a full-orbit phase curve measurement shows symmetric phase variation with no evidence for day-night heat redistribution, pointing towards the bare rock interpretation \citep{zhang24}.% While the analyses for all three planets conclude that they are most likely airless, thin atmospheres  without strong absorption features can also fit the data.  

For the other planets in the sample, TRAPPIST-1b and c and LHS 1478b, the picture is somewhat more complex. TRAPPIST-1b is consistent with a zero-albedo bare rock at 15~$\mu$m, but cooler at 12.8~$\mu$m \citep{greene23,ducrot24}. The measurements could be explained by an airless planet with a surface temperature intermediate between the two measured values. Alternatively, there could be an atmosphere with a \ce{CO2} \emph{emission} feature at 15 $\mu$m, in an atmosphere with a thermal inversion caused by photochemical hazes. This scenario can reproduce the higher temperatures at 12.8~$\mu$m but requires some fine-tuning of the haze properties. For TRAPPIST-1c, the results are similarly degenerate. The $15\mu$m brightness temperature is intermediate between zero and full heat redistribution. Several atmosphere models are consistent with the data, including steam compositions and up to 10 bars of \ce{O2} with trace amounts of \ce{CO2} \citep{lincowski23}. However, realistic \ce{CO2} atmospheres like that of Venus are disfavored (at 3.1$\sigma$ for cloud-free compositions and 2.6$\sigma$ for H$_2$SO$_4$ clouds).  To resolve the degenerate interpretations for TRAPPIST-1b and c, additional data are forthcoming; a joint phase curve for both planets was recently observed at 15~$\mu$m to measure the day-night heat redistribution and definitively test whether an atmosphere is present on either one (GO 3077; PI: Gillon). Finally, LHS~1478b is also an ambiguous case; there is tentative evidence for atmospheric heat redistribution, but the data are affected by instrument systematics that make a robust secondary eclipse measurement challenging \citep{august25}.

To conclude our summary of rocky planet thermal emission measurements, we briefly touch on 55 Cancri e. This planet is very different from the M-dwarf planets described above; it orbits a Sun-like star and has a much hotter equilibrium temperature, nearly 2000 K \citep{winn11}. It is right on the border of what could be considered a rocky planet, with a density consistent with pure silicate and a size of 1.9 Earth radii, consistent with a Ca/Al-rich interior or dissolved volatiles in a magma ocean \citep{crida18,dorn19,kite21,dorn21}. The first JWST observation of 55 Cancri e revealed an absorption feature consistent with either CO or \ce{CO2}, suggesting a thick volatile-rich atmosphere sustained by outgassing from a magma ocean below \citep{hu24}. However, four subsequent observations did not confirm the first measurement, but instead revealed a highly variable brightness temperature at 4.5$\mu$m from 900 to 2300 K \citep{patel24}. Such variability could be produced by sporadic outgassing, or by circumstellar dust. It could also be due to uncorrected instrument systematics in the data; 55 Cancri e is a very bright star, and pushes the JWST instruments to their limits. Clearly, more work is needed before we have the final word on 55 Cancri e, but for now, the debatable hints of atmospheric absorption features do agree with predictions from interior structure models that a substantial atmosphere is present \citep{crida18}.

\section{Discussion and Future Prospects}
In its first few years, JWST has achieved several observational milestones for rocky exoplanets, including the most precise transmission spectra to date and the first thermal emission detection for planets below 800 Kelvin. Nevertheless, despite these breakthroughs, JWST still has not definitively detected an atmosphere on a rocky planet. This is not a huge surprise, as the first programs focused on low-hanging fruit that is the easiest to observe, and many plausible physical scenarios have not yet been explored.

The most precise JWST transit observations have now reached the sensitivity required to detect cloud-free, water-rich atmospheres (with $\mu =18$  g/mol). While this is a big step forward compared to previous results that could only detect extended, hydrogen-rich atmospheres \citep{wordsworth22}, many possible compositions have features that are still below the noise level of the current data. This includes atmospheres dominated by  \ce{N2}, \ce{O2}, and \ce{CO2} ($\mu$ = 28, 32, and 44 g/mol, respectively). These heavier molecules decrease the scale height of the atmosphere and shrink the amplitude of features in the spectra. At the current level of precision, some of the measured spectra do show possible evidence of absorption features, but these are on the edge of significance and some may be affected by stellar contamination \citep[e.g.][]{MayMacDonald2023,MoranStevenson2023,Cadieux2024b,Damiano2024, Gressier2024}. Definitive detections, for a broader range of atmospheric scenarios, will require additional data.

To push the sensitivity of future observations, we suggest  a ``five scale height challenge": transmission spectra should reach the precision necessary to confidently detect a feature with an amplitude of five scale heights in a nitrogen-rich atmosphere, analogous to the \ce{CO2} feature in Earth's spectrum (see Figure~\ref{fig:earth}). Nitrogen-rich atmospheres are the next milestone after water-rich atmospheres, and making this step will open up discovery space to a wider range of plausible compositions, and solidify (or refute) the hints of spectral features identified in current data. Note that this general guideline does not account for all the possible effects shaping the feature size, including aerosols and temperature structure. Rather, it represents a simple step up in precision from the current results that are mainly sensitive to features in water-rich atmospheres (with 60\% larger scale height than that of an N$_2$-dominated composition). Reaching a precision of five scale heights in a nitrogen-rich atmosphere is an ambitious, but not impossible, goal --- as noted in \autoref{sec:transit}, the most precise spectrum measured to date, for GJ~486b, already has an average uncertainty of 3.8 $N_2$ scale heights at a resolution of 50 nm. Given JWST's excellent performance, with close to photon-limited precision and no clear evidence for an instrumental noise floor, chances are good that this challenge can be met for the highest S/N targets, at least from an instrumental standpoint.

The biggest roadblock that has emerged for precise transmission spectroscopy is stellar contamination. This effect is strongest for planets transiting mid-to-late M-dwarf stars (which are rich in molecular features) so the rocky planets with the best planet-to-star radius ratios will be particularly affected \citep{rackham18}.  Indeed, several of the JWST spectra show strong evidence for contamination that varies over time \citep{Lim23,Cadieux2024b}.  For the most extreme case (i.e., the fast-rotating, late-M star, TRAPPIST-1), the stellar contamination correction is already the dominant source of uncertainty in the spectrum \citep{Lim23}.  This effect appears to be less problematic for slow-rotating, early-to-mid type M dwarfs.  Ultimately, stellar contamination may be the limiting factor in the achievable precision, and will need careful attention in coming years \citep{rackham23}. Some creative approaches are already emerging, such as using back-to-back transit measurements in multi-planet systems to empirically correct for contamination \citep{rathcke25}.

As with the transmission spectra, none of the thermal emission measurements have shown unambiguous evidence of atmospheres. By the principle of Occam's razor, the simplest explanation that fits the data is that all of the planets observed so far are bare rocks, with negligible heat redistribution. In this case, the mean measured brightness temperature of $0.95\pm0.01 \times T_\mathrm{B,max}$ corresponds to a (wavelength-independent) Bond albedo of $A_B = 0.15\pm0.03$. This is broadly consistent with predictions for plausible, highly absorptive rocks such as basalt. However, an important caveat is that a one-to-one mapping between albedo and composition is complicated by degeneracy between different surface textures and also the wavelength dependence of the albedo \citep{hu2012surface,fortin24,paragas25}.  In addition, Earth-like atmospheres can also have low Bond albedos, though they would show signatures of atmospheric heat redistribution and molecular features that have not been seen in the data \citep{kopparapu13,rugheimer15, madden20}.

In addition to bare rock spectra, there are some model atmospheres that are compatible with the data. The main requirement is that the atmospheres are not too thick (surfaces pressures typically $\lesssim 10$ bar). \ce{CO2}-rich compositions are also disfavored due to their strong absorption features, particularly at 15 $\mu$m. While many atmospheres fit these criteria, Venus-like atmospheres appear to be rare, at least for the relatively hot, close-in planets observed so far.  Further work is needed to explore the effects of realistic clouds in Venus-like atmospheres; however, for the specific case of TRAPPIST-1c, thick, CO$_2$-rich atmospheres are still disfavored even when sulfuric acid clouds are included \citep{lincowski23}. 

The absence of thick, \ce{CO2}-rich atmospheres on hot rocky planets is already an intriguing result. For planets with a large initial volatile inventory, tens of bars of \ce{CO2} could  accumulate via volcanic outgassing, similar to Venus \citep[e.g.][]{dorn18, foley18, bower19}. Substantial \ce{O2} can also accumulate through desiccation, though the abundance can be capped by the combustion-explosion limit \citep[e.g.][]{luger15, grenfell18}.  \ce{CO2} is particularly resistant to atmospheric loss due to its higher mean molecular weight and low susceptibility to photodissociation. Indeed, models predict that \ce{CO2}-dominated atmospheres should be stable for Gyr-timescales, even at high levels of irradiation \citep{tian09}.  The fact that thick, \ce{CO2}-rich atmospheres have \emph{not} been seen in emission suggests that a limited reservoir of carbon and oxygen were present at the planets' formation, or that these species were lost very early in the planets lifetime, dragged along by hydrodynamically escaping hydrogen \citep{teixeira24}. It is unlikely that substantial volatiles persist in the planets' interiors; while \ce{H2O} and \ce{CO2} are highly soluble in magma \citep{dorn21}, these species rapidly outgas as the mantle solidifies and cools \citep{bower19} unless a very thick volatile envelope ($>100$ bar) is present to warm the surface \citep{kite21}. Such thick envelopes are disfavored by the low heat redistribution observed for this sample.

If the planets considered here are strongly affected by atmospheric escape, our best chances of detecting atmospheres may be for cooler worlds  -- planets on the right side of the cosmic shoreline (\autoref{fig:shoreline}). While highly irradiated rocky planets are expected to suffer substantial atmospheric loss due to atmospheric escape, particularly those with M-dwarf hosts, more temperate planets can more easily maintain substantial atmospheres \citep{luger15,bolmont17}. If atmospheric loss is XUV-driven, earlier type host stars with relatively less XUV flux relative to bolometric luminosity would also be more favorable for atmosphere retention \citep{france16}.

In summary, JWST has made rapid progress on rocky planet atmospheres in its first few years of science operations. Looking ahead, there is a strong push from the exoplanet community to continue work on this topic. This includes a recently approved 500-hour Director's Discretionary Time program, ``Rocky Worlds," that is dedicated to observing thermal emission from up to a dozen planets, including some with equilibrium temperatures below 500 K. In addition to the DDT program, dozens of General Observer programs are underway, ranging from characterizing the surface composition of airless worlds to studying possible rock vapor atmospheres on lava planets. With an expected mission lifetime greater than ten years, JWST has ample time to push towards some of the challenges outlined in this article --- higher precision transmission spectra and thermal emission for cooler targets. If there are atmospheres to be found on rocky, M-dwarf exoplanets, JWST can and will find them. Either way, the observatory is certain to leave a lasting legacy in rocky planet characterization.

\matmethods{The data in this work come from previously published JWST transmission and emission spectroscopy measurements for rocky exoplanets. For display purposes, we binned the transmission spectra in Fig. 3 to a common wavelength scale. The binned spectra are available in the Zenodo repository \url{https://doi.org/10.5281/zenodo.15084225}. For the thermal emission measurements, we normalized the dayside brightness temperature relative to that of a bare rock with zero heat redistribution and zero albedo, using Equation 6 from \cite{mansfield19}.}

\showmatmethods{} % Display the Materials and Methods section

\acknow{
% Please include your acknowledgments here, set in a single paragraph. Please do not include any acknowledgments in the Supporting Information, or anywhere else in the manuscript.
KBS acknowledges support from the National Aeronautics and Space Administration (NASA) under Grant No. 80NSSC23K1399 issued through the Interdisciplinary Consortia for Astrobiology Research (ICAR) program. The authors thank Matej Malik for sharing the pure \ce{CO2} atmosphere models from \citep{malik19}.
This research has made use of the NASA Exoplanet Archive, which is operated by the California Institute of Technology, under contract with the National Aeronautics and Space Administration under the Exoplanet Exploration Program.
}

\showacknow{} % Display the acknowledgments section

\bibsplit[2]
%Use \bibsplit to split the references from the body of the text. Value "[2]" represents the number of reference in the left column (Note: Please avoid single column figures & tables on this page.)

% Bibliography
\bibliography{pnas-sample}

\end{document}